\documentclass{ws-ijmpd}
\usepackage{amsmath}
\usepackage{latexsym}
\usepackage{amsfonts}
\usepackage{bm}
\usepackage{mathrsfs}
\usepackage{amssymb}
\begin{document}

\markboth{Yi Zhang,Yungui Gong,Zong-Hong Zhu}
{modified gravity emerging from thermodynamics and holographic principle}

%
\catchline{}{}{}{}{}
%

\title{MODIFIED GRAVITY EMERGING FROM THERMODYNAMICS  AND HOLOGRAPHIC PRINCIPLE  }

\author{YI ZHANG}

\address{College of Mathematics and Physics, Chongqing University of Posts and Telecommunications,\\
Chongqing 400065, China\\Department of Astronomy, Beijing Normal university,\\
Beijing 100875, China\\
zhangyia@cqupt.edu.cn}

\author{YUNGUI GONG}

\address{College of Mathematics and Physics, Chongqing University of Posts and Telecommunications,\\
Chongqing 400065, China\\
gongyg@cqupt.edu.cn}

\author{ZONG-HONG ZHU}

\address{Department of Astronomy, Beijing Normal university,\\
Beijing 100875, China\\
zhuzh@bnu.edu.cn}

\maketitle

\begin{history}
\received{Day Month Year}
\revised{Day Month Year}
\comby{Managing Editor}
\end{history}

\begin{abstract}
 A new conception is proposed  in Ref.\cite{Verlinde:2010hp,Padmanabhan:2009kr} that gravity is  one kind of entropic force. In this letter,
we try to discuss its applications to the modified gravities by using three different corrections
to the area law of entropy which are derived from the quantum
effects and extra dimensions. According to the assumption of
holographic principle, the number of bits $N$
which is related to the equipartition law of energy is modified.
Then, the modified law of Newton's gravity and the modified Friedmann equations are obtained by using the new
notion.   By
choosing suitable parameters, the modified area law of entropy leads
to de-Sitter solutions which can be used to explain the accelerating
expansion of our universe. It  suggests that the accelerating
phase in our universe may be an emergent phenomenon based on
holographic principle and thermodynamics.
\end{abstract}


\section{Introduction}\label{sec1}
The  gravity was always being regarded as one kind of fundamental  force.  However, the origin of gravity is still an unaccessible theoretic puzzle. Recently,
a new conception of gravity is proposed by Ref.\cite{Verlinde:2010hp,Padmanabhan:2009kr}. The fundamental basement of these work could be traced back to   the profound connections between gravitation and thermodynamics which are suggested by the discovery of the black hole entropy \cite{Bekenstein:1973ur}, the four laws of classical black hole mechanism \cite{Bardeen4} and the Hawking radiation \cite{Haw}.  Based on
thermodynamics and holographic principle, the new idea, that the gravity can be explained as one kind of entropic force, gives out the Newton's law. But from a point of view of thermodynamics, is it possible to derive the Einstein equations of gravitational fields? Jacobson first answered
this question  based on the geometric feature of thermodynamic
quantities of black holes whose result is yes \cite{Jac}.
Referred to  the
new idea of  entropic force about gravity, the Friedmann
equation which is the time-time component of Einstein equation could be gotten
as well \cite{Shu:2010nv,Cai:2010hk}.

Now, the gravity has the new explanation that it can be treated as one
kind of entropic force. In this new theoretic frame, what about
the  observable effects? Since cosmology covers all energy scales
in our universe, our universe could be regarded as a natural lab
to detect the observable effect especially for the  gravitational
force. One of the most important development in cosmology is the
proposal of the two accelerations in the early and late times. Inflation,
which is an accelerating phase in the early time, was introduced as a way
to solve the problems in the standard big-bang theory \cite{Guth:1980zm,Linde:1981mu}.
Meanwhile, the current observable accelerating expansion of our universe was suggested by combining
different cosmic probes that primarily involves Supernova data \cite{Spergel:2003cb,SN1}.
The simplest candidate of the two accelerations is the $\Lambda CDM$ model which has the
coincidence problem. Nevertheless, the validity of General
Relativity on large astrophysical and cosmological scales has never been tested but only
assumed, the acceleration in early and late universe might be  signals of
a breakdown of General Relativity
\cite{Briscese:2007cd,Hu:2007pj,Starobinsky:2007hu,Appleby:2007vb}. And, the breakdown of
General Relativity is fairly working at solar system and in the weak field regime \cite{Horowitz:1980fj,Hayashi:1980av}.

Based on   the Bekenstein-Hawking entropy in General Relativity \cite{Haw,Modesto19}, the entropy
of a Schwarzchild black hole    is used in Verlinde's work,
 \begin{equation}
 \label{S}
 S=\frac{k_B c^{3}}{4G\hbar}A,
 \end{equation}
where $G$ is Newton's constant,  $A$ is the surface area of the
horizon; and it is the standard relation between horizon entropy and
horizon area. Phenomenologically, one could assume
that the horizon entropy is a general function of the horizon area,
\begin{equation}
S=\frac{k_B}{4} \,g(\frac{ c^{3}A}{G\hbar}),
\end{equation}
where $g$ is an arbitrary function. At least, after taking quantum correction into account, the form of the entropy in the modified gravity will be changed. However, current observations could not tell us the law of gravity at large or small scales. If we start from the corrected relation between the entropy and the area, can we get the modified Einstein equations? Does this modified gravity make our universe accelerate?

To answer the above questions,  in this letter, we try to derive the law of gravity with the modification of area law of
entropy by the new notion of gravity which is treated as entropic force.  Specifically, we discuss three different
corrections for the area law of the entropy.  One is the most
popular used logarithmic correction which is derived from quantum
effects \cite{Cai:2008ys,Smolin20,Lidsey22,Fursaev:1994te,Banerjee:2008fz,32,12,0,positive,loop}(e.g. the conformal anomaly \cite{Cai:2008ys}).  One is
the power-law correction which is related to the extra dimensions
\cite{Wang:2005bi,turner,dgp,lisa,gong09,braneworld}. The last one is the
$f(R)$ correction \cite{noe}  which  may break the energy
equipartition law \cite{fR,Eling:2006aw,Bamba:2009id}.  Besides the modified gravities, the dark energy model from entropic force has been discussed as well \cite{Li:2010cj,Gao:2010fw,Wei:2010am}.

We arrange our paper as follows. In Sec.\ref{sec2}, we give out the
approach of   getting the  Newton's law by using Verlinde's
method. Next,  we introduce the method of getting the
 Friedmann equations with the energy equipartition law in Sec.\ref{sec3}.
Specifically, in Sec. \ref{sec4}, we discuss the logarithmic
correction which may be related to the quantum effects. Then, we discuss the
power-law correction which gives an accelerating universe in Sec.
\ref{sec5} and we consider the $f(R)$ correction in Sec.
\ref{sec6}. Finally, the paper is concluded in Sec. \ref{sec7}.

\section{The Newton law}\label{sec2}
 The essence of Verlinde's idea is  that
gravity is not fundamental; it is a kind of  entropic
force. The process of deriving the Newton's law mainly depends on thermodynamics and holographic principle. The notion of the entropic force can be expressed as
 \begin{equation}
\label{ef}
 F\triangle x=T\triangle S,
 \end{equation}
where $\Delta S$ is the change of entropy of the gravitational
system, $\Delta x$ is the displacement of a particle in the
gravitational  system, and $T$ is the temperature of the system.
Motivated by Bekenstein's original argument on black holes
\cite{Bekenstein:1973ur}, it is postulated that since the change
of entropy $\Delta S$ is associated with the information on the
boundary and it equals $2\pi k_B$ when $\Delta x=\hbar/m$, the change of
entropy is assumed as the following form
 \begin{equation}
 \label{a1} \Delta S= 2\pi k_B {mc\over\hbar} \Delta x,
 \end{equation}
 where $k_B$ is the  Boltzmann's  constant, $c$ is the velocity of light, and $\hbar$  is the reduced Planck constant.

 The holographic principle assumes that all the information about black
holes is encoded in the surface; it could be used in the Schwarzschild black hole and the de-Sitter space.
If the number of bits $N$ of the holographic system is proportional
to the area of the holographic screen,  the following relation will be generalized  by
the holographic principle
 \begin{equation}
 \label{SN}
 N=\frac{4S}{k_{B}}.
  \end{equation}
 Specially,  based on the entropy  in general relativity as Eq.(\ref{S}) expressed, the  above equation can be rewritten as
 \begin{equation}
 \label{bits} N =\frac{Ac^{3}}{ G\hbar}=\frac{4\pi r^2c^{3}}{ G\hbar},
 \end{equation}
 where the radius of gravitational system is $r$ and the area of the corresponding holographic screen is $A=4\pi r^{2}$.

One of the most important assumptions for the notion of gravity of entropic force is  that each bit on the holographic screen
contributes an energy of $k_B T/2$ to the system.  The equipartition law of energy in thermodynamics can be   used
 \begin{equation} \label{E}
 E=\frac{1}{2}Nk_BT=\frac{2\pi c^{3} r^2}{G\hbar}k_BT=Mc^{2},
 \end{equation}
 where the energy is  defined. Amazingly,
combined Eqs. (\ref{ef}), (\ref{a1}) and (\ref{E}), the  Newton's
Law of gravitation is derived,
 \begin{equation}
 \label{8}
  F=G\frac{Mm}{r^{2}}.
   \end{equation}
The above calculation depend on  the holographic
principle and the energy equipartition law closely. It is interesting to extend Verlinde's idea from general relativity to the more generalized gravity.

\section{The Einstein  Equations}\label{sec3}
The research on  Einstein equation
is a useful approach to investigate the dynamics of our universe.
And, we focus on a $(3+1)$-dimensional flat
Friedmann-Robertson-Walker (FRW) universe with the metric in
double null form,
\begin{equation}
 \label{metric}
 ds^2 = h_{ab}dx^adx^b +\tilde r^2 d\Omega_{2}^2,
 \end{equation}
 where $a$ is the scale factor,  $\tilde r = a r$, the
 $2$-dimensional metric is $h_{ab} = {\rm diag} (-1, a^2)$ and the unit spherical metric is given by $d\Omega_{2}^{2}=d\theta^{2}+\sin^{2}\theta d\varphi ^{2}$.
 The energy momentum tensor $T_{\mu \nu}$ of  matter
in the universe is supposed to  have the perfect fluid form
 \begin{equation}
 T_{\mu \nu}=(\rho
+p)U_{\mu}U_{\nu} +p g_{\mu\nu},
 \end{equation}
 where $\rho$ and $p$ are the
energy density of matter and the pressure of matter separately. The evolution of
energy density for matter could be assumed as,
 \begin{equation}
 \label{con}
 \dot{\rho}+3H(\rho+p)=0,
 \end{equation}
  where $H\equiv
 \dot a /a$ denotes the Hubble parameter. The derivation of the Einstein equations will mainly depend on two different definitions of the energy, one is from the energy equipartition law which is related to the notion of gravity as entropic force, the other one is from the Misner-Sharp mass. In the next sections, we will introduce them separately.

\subsection{ The Energy Equipartition  Law }

The notion of  horizon  plays an important role in cosmology.
The apparent horizon, the event horizon  and the particle horizon are
different both in definitions and physical meanings. In cosmology,
the proper
chosen horizon   is just the apparent horizon which coincides with the double null metric (\ref{metric}) and it is
the boundary surface of anti-trapped region \cite{Cai:2008gw}.
The apparent horizon
 is defined as $h^{ab} \partial_a \tilde r
 \partial _b \tilde r=0$, it turns out to be
\begin{equation}
 \label{rA}
 \tilde r_A = \frac{c}{H}.
 \end{equation}
 Only in the flat FRW metric, the above form of apparent horizon is the same as the Hubble horizon.   From the definition of the
apparent horizon, we get
\begin{equation}
\label{dra} d\tilde{r}_A=-\frac{\tilde{r}_A^3}{c^{2}}H\dot{H}dt.
\end{equation}
During the time interval $dt$, the radius of the apparent
horizon is supposed to evolve from $\tilde{r}_A$ to $\tilde{r}_A+d\tilde{r}_A$,
then the change of the area of the apparent horizon is
$dA=8\pi\tilde{r}_A d\tilde{r}_A$. Here, we try to focus on the accelerating phases where the spacetime is quasi-static and the holographic conjecture is reasonably held.

Furthermore, the Hawking temperature  corresponding the horizon  is
 \begin{equation}
 \label{TA} T=\frac{c\hbar}{2\pi k_ B\tilde{r}_A}.
 \end{equation}
And, the
change of the horizon temperature is
\begin{equation}
\label{dTA} dT=-\frac{c\hbar}{2\pi k_B\tilde{r}_A^2}d\tilde{r}_A.
\end{equation}
The Hawking temperature is determined by geometry itself and it has
nothing to do with gravity theory explicitly. Once the
geometry is given, the surface gravity and the Hawking temperature
can be determined immediately.

Since the apparent horizon with area $A=4\pi \tilde{r}_A^2$
carries $N$ bits of information, from  Eq.(\ref{SN}), we could get the
change of the number of bits $dN$. Then, from the energy
equipartition rule, we  get the changes of the total energy is gotten,
\begin{equation}
\label{dE} dE=\frac{k_B}{2}NdT+\frac{k_B}{2}T dN.
\end{equation}
The expression of $dE$ is determined by the geometry with
variables such as the Hubble parameter $H$. It is  the definition of energy from the view point of the thermodynamics.

\subsection{ The Misner-Sharp Mass }
On the other side, for a spherically symmetric space-time with the
metric (\ref{metric}), using the Misner-Sharp mass \cite{Misner:1964je,CK,Hayward2,Hayward3,Bak:1999hd}
\begin{equation}
{\cal{M}}=\tilde{r}(1 - g^{ab}\tilde{r}_{,a}\tilde{r}_{,b})/2G,
 \end{equation}
the $a-b$
components of the Einstein equation give the mass formulas
\cite{Gong:2007md,Gong:2006ma}
\begin{equation}
\label{msmasseq} {\cal{M}}_{,a}=4\pi\tilde{r}^2(T_a^b-\delta_a^b
T)\tilde{r}_{,b},
\end{equation}
where $T=T^a_a$.
At the apparent horizon, the Misner-Sharp mass ${\cal
M}=4\pi\tilde{r}_A^3\rho/3$ can be interpreted as the total energy
inside the apparent horizon. To project the mass formulas, one could
use the  generator $k^a=(-1,\ Hr)$ of the apparent
horizon, which is null at the horizon. Since $k^a\tilde{r}_{,a}=0$,
using the mass formulae (\ref{msmasseq}),  the energy which flows
through the apparent horizon could be gotten
 \begin{eqnarray} \label{dE2}
\nonumber d E &=&k^a\nabla_a {\cal M} dt=4\pi \tilde{r}^2 T_a^b\tilde{r}_{,b}k^adt \\
&=& 4\pi\tilde{r}_A^3H(\rho+p)dt.
 \end{eqnarray}
 This is the other definition of the energy which is related to the matter part.

\subsection{The Equations In Cosmology}
Combined with Eqs.(\ref{dE}) and (\ref{dE2}), the so-called Raychaudhuri equation   is gotten,
\begin{equation}
\label{20}
\frac{k_B}{2}NdT+\frac{k_B}{2}T dN=4\pi\tilde{r}_A^3H(\rho+p)dt,
\end{equation}
which connects the geometry and the matter.
 Considering the evolution of matter, after  integration,  the Friedmann equation which is the time-time component of Einstein equations could be derived easily. And, the other component of Einstein equations can be derived by combining the Raychaudhuri equtaion and the Friedmann equation.

In a summary, the Misner-Sharp mass related to the Einstein equations decides the right side of Eq.(\ref{20}). Meanwhile, the left side of Eq.(\ref{20}) is decided by the equipartition law of energy which is related to the thermal dynamical system. The Misner-Sharp mass which is derived from the Einstein equation is just a variable. To make a correct Raychaudhuri equation, both sides of Eq.(20) must be correct. Because of the validity of the Misner-Sharp mass, we only need to consider the validation of the equipartition law of energy. There is a question that whether we can apply this method to the modified area law of entropy, explicitly speaking, whether the energy equipartition law can be used in modified Einstein gravity.  Firstly,  the assumption,  that the apparent horizon has
an entropy proportional to its horizon area, is originated from the
black hole thermodynamics which  obeys the so-called area formula.
The modified theory of gravity may explain the present acceleration without introducing dark energy. If it so happens, Einstein gravity will be only an approximate theory, and the relation of the entropy to the horizon's area  should be changed. Therefore, it is interesting to see thermodynamic properties of the universe may be modified. Secondly, the energy equipartition law only exists in the thermal equilibrium processes. If we can get the correct Raychaudhuri equation (or the correct Friedmann equation), the using of the equipartition law of energy is proper. But if we can not get the correct Friedmann equation, the using of the equipartition law of energy is problematic.

\section{The Logarithmic Correction}\label{sec4}

 The logarithmic correction to the
horizon entropy can  arise from the breaking of  the
conformal invariance of classical massless fields or an
anomalous trace for the energy-momentum tensor. As a result, the
entropy becomes \cite{Cai:2008ys}
 \begin{equation}
 \label{39}
  S_1=\frac{k_{B}c^{3}}{4G
\hbar}A+ \frac{\alpha k_{B}}{4}\ln \frac{Ac^{3}}{ G\hbar},
 \end{equation}
where  the subscript ``$1$" notifies  the logarithmic correction for
entropy, $\alpha$ is a dimensionless parameter whose value is still in
debate because there are other physical origins about the  logarithmic
  correction, e.g. such a term appears in the study of black
  hole entropy of the loop quantum gravity \cite{32,12,0,positive,loop} and in  the
  correction to black hole entropy of the thermal equilibrium
  fluctuation or the quantum fluctuation \cite{flu}. Therefore, there are different values of $\alpha$, $\alpha=-3/2$,  $\alpha=-1/2$, $\alpha=0$ and $\alpha>0$ are given   in Refs.\cite{32,12,0,positive} separately.
Based on the holographic principle,
the number of bits which is corresponding the quantum corrected
entropy-area relation then becomes
 \begin{equation}
 \label{bits1}
  N_{1}=\frac{A c^{3}}{ G\hbar} + \alpha \ln \frac{  Ac^{3}}{ G\hbar}.
 \end{equation}
Using the method in Sec.\ref{sec2} and Sec.\ref{sec3},  the Newton's law and the Friedmann equation of General Relativity could be extended to the gravity with a logarithmic correction.

\subsection{The Modified Newton's Law For Logarithmic Correction}
By substituting the modified number of bits $N_1$ into the
equipartition law of energy (\ref{E}), the modified
energy of the system is gotten. Then  combining this modified energy with
Eqs.(\ref{ef}) and (\ref{a1}), the Newton's law could be extended to the gravity with a logarithmic correction,
 \begin{equation}
  F_1=\frac{G}{1+\alpha \ln (\frac{ c^{3}A}{G\hbar})/(\frac{ c^{3}A}{G\hbar})}\frac{M m}{ r^{2}}=
  \frac{G}{1+\alpha \ln (\frac{4 \pi r^{2} c^{3}}{G\hbar})/(\frac{\pi r^{2} c^{3}}{G\hbar})}\frac{M m}{ r^{2}}.
 \end{equation}
 It is the Newton's law for the logarithmic correction of gravity. Here, our discussions of the modified Newtonian forces are based on the entropic force which  is a quality of thermal statistics. After choosing the proper parameter, it could get back to the Newtonian gravity which satisfies the observations.
For example,
 when $ \alpha \ln (\frac{4\pi r^{2} c^{3}}{G\hbar})/(\frac{4\pi r^{2} c^{3}}{G\hbar})\ll1$, the corresponding leading terms of the corresponding Taylor
 expansion are
 \begin{equation}
 \label{new1}
 F_1=\frac{G M m}{
r^{2}}\left[1-\alpha \ln (\frac{4\pi
r^{2}c^{3}}{G\hbar})/(\frac{4\pi r^{2}c^{3}}{G\hbar})+\left(\alpha
\ln (\frac{4\pi r^{2}c^{3}}{G\hbar})/(\frac{4\pi
r^{2}c^{3}}{G\hbar})\right)^{2}\right].
 \end{equation}
 In the next section, we try to discuss the modification of Friedmann equation due to the correction of the law of gravity.

\subsection{The Friedmann Equation For Logarithmic Correction}

For the gravity of the logarithmic correction, the corresponding thermodynamics is changed.
Since the apparent horizon with area $A=4\pi \tilde{r}_A^2$ carries
$N_1$ bits of information, then from  Eq.(\ref{bits1}), one could  get the
change of the number of bits
 \begin{equation}
 \label{dbits1} d N_1 =\frac{8\pi c^{3} }{ G\hbar} \tilde{r}_{A}d \tilde{r}_A+
 2\alpha \frac{d \tilde{r}_{A}}{ \tilde{r}_{A}}.
\end{equation}
 From the energy equipartition rule, one then obtains the changes of the
total energy
\begin{equation}
\label{dE1} dE_1=\frac{c^{4}}{G}\left(1+\alpha\frac{ G \hbar}{2\pi
c^{3}\tilde{r}_A^{2}}-\alpha\ln (\frac{4\pi
\tilde{r}_A^{2}c^{3}}{G\hbar})/(\frac{4\pi
\tilde{r}_A^{2}c^{3}}{G\hbar})\right)d\tilde{r}_A.
\end{equation}
Combined with the definition of the apparent horizon, the above equation becomes
\begin{equation}
\label{dE11} dE_1=\frac{c^{4}}{G}\left(1+\alpha \frac{ G \hbar
H^{2}}{2\pi c^{5}} -\alpha \ln(\frac{ 4\pi c^{5}}{G \hbar H^{2}}
)/(\frac{ 4\pi c^{5}}{G \hbar H^{2}})\right)d\tilde{r}_A.
\end{equation}
On the other side, the Misner-Sharp mass $ dM_1=-4\pi \tilde{r}_A^3(\rho+p)c^{2}Hdt$ is decided by the matter part. Based on the two definitions of  energies for the logarithmic correction,
 the following equation is obtained
 \begin{equation}
 \label{FE1}
 \dot H \left(1+\alpha \frac{ G \hbar
H^{2}}{2\pi c^{5}} -\alpha \ln(\frac{ 4\pi c^{5}}{G \hbar H^{2}}
)/(\frac{ 4\pi c^{5}}{G \hbar H^{2}})\right)=-4\pi G (\rho +p).
 \end{equation}
Using the energy conservation equation  of matter (\ref{con}), one then has the
Friedmann equation
  \begin{equation}
  \label{h21}
  H^2 +\frac{3\alpha G\hbar}{16\pi c^{5} }  H^{4}+\frac{\alpha G\hbar H^{4}}{8\pi c^{5}}\ln (\frac{ G\hbar H^{2}}{4\pi c^{5}})= \frac{8\pi G}{3}\rho+\Lambda_1,
  \end{equation}
where $\Lambda_1$ is the integral constant
which can be regarded as a cosmological constant.
 Obviously,
 a de-Sitter phase can be gotten  when $\frac{8\pi
G}{3}\rho\ll\Lambda_1$.
 And when $\alpha<0$,  we can also get de-Sitter solutions
as the the accelerating phase.  The de-Sitter solution  with the  logarithmic
correction is in the  quasi-static spacetime where the holographic principle makes sense.
 Furthermore,  if we neglect the term $ \ln
(\frac{4\pi \tilde{r}_A^{2} c^{3}}{G\hbar})/(\frac{4\pi
\tilde{r}_A^{2} c^{3}}{G\hbar})$, one could get
\begin{equation}
  H^2 +\frac{\alpha G\hbar}{4\pi c^{5} }  H^{4}\simeq\frac{8\pi G}{3}\rho+\Lambda_1.
  \end{equation}
  The $H^{4}$ term still exists, while the Newton's law (\ref{new1}) is satisfied at the same time.

In  a short summary, for the logarithmic correction of  gravity, the field equation has additional terms, e.g. the $H^{4}$ term. With proper parameters, the Newton's law could be recovered. When the parameter $\alpha$ is negative,  the accelerating phase could be gotten.

\section{The Power-law Correction}\label{sec5}
Extra
dimensional theories with infinite volume  modify the law of gravity
in the far infrared energy scale (large distance is called as well). At small distances, its gravitational
dynamics is very close to the standard $4$-dimensional Einstein
gravity \cite{dgp}. Theories with large extra dimensions are  motivated by both the
hierarchy and cosmological constant problems
 \cite{dgp,lisa}.  Corresponding to the theories with large extra dimension(s), we choose the power-law
correction to the horizon entropy for extra dimension which  could be
expressed as \cite{Wang:2005bi}
\begin{equation}
S_{2}=k_B\frac{ Ac^{3}}{4 G\hbar}(1-KA^{\beta}),
\end{equation}
where $K$ is a constant parameter, the subscript ``$2$" means the
power-law correction background. Correspondingly, the number of bits
could be written as
 \begin{equation} N_{2}=\frac{Ac^{3}}{ G
\hbar}(1-r_{c}^{-2\beta}A^{\beta}),
 \end{equation}
where $r_{c}^{-2\beta}=4K /k_B$ has the distance's dimension. The modified number of bits $N_{2}$ will change the thermodynamics as well.

\subsection{The Modified Newton's Law For The Power-law Correction}

With this definition of the number of bits $N_2$, using
Eqs.(\ref{ef}), (\ref{a1}) and (\ref{E}), the modified
law of gravity could be easily gotten,
  \begin{equation}
F_2=\frac{G}{1-(\frac{4\pi r^{2}}{r_{c}^{2}})^{\beta}}\frac{M m}{
r^{2}}.
 \end{equation}
When $(\frac{4\pi r^{2}}{r_{c}^{2}})^{\beta}\ll1$, the related leading terms of the
Taylor expansion could be obtained
 \begin{equation}
 F_2=\frac{G M m}{
r^{2}}\left(1+(\frac{4\pi r^{2}}{r_{c}^{2}})^{\beta}-(\frac{4\pi
r^{2}}{r_{c}^{2}})^{2\beta}\right).
 \end{equation}
The cutoff parameter $r_{c}$ is critical for the dynamics. When $\beta>0$, in  small scales where $r\ll r_{c}$, the above
equation is reduced to Newton's gravity.

\subsection{The Modified Friedmann Equation For The Power-law Correction}
For the power-law correction of the entropy, the change of $N_2$ is
 \begin{equation}
 dN_{2}=\frac{8\pi c^{3}\tilde{r}_{A}}{ G
\hbar}\left(1-(1+\beta)(\frac{4\pi
\tilde{r}_A^{2}}{r_{c}^{2}})^{\beta}\right)d \tilde{r}_{A}.
 \end{equation}
Then the change of the energy defined by the energy equipartition rule is
 \begin{equation} \label{DE22}
dE_{2}=\left(1-(1+2\beta)( \frac{4\pi
\tilde{r}_A^{2}}{r_{c}^{2}})^{\beta}\right)\frac{c^{4}}{G}d\tilde{r}_A.
 \end{equation}
Meanwhile, the Misner-Sharp mass is decided by  matter. Therefore, using Eqs. (\ref{dE2}) and (\ref{DE22}),
 the Raychaudhuri equation could be obtained
\begin{equation}
\dot{H}\left(1-(4\pi)^{\beta}(1+2\beta)(\frac{c}{r_{c}H})^{2\beta}\right)=-4\pi
G (\rho+p).
 \end{equation}
  Combining the energy conservation equation, one could
get the modified Friedmann equation in the power-law correction as well,
   \begin{equation}
  H^2 \left(1-\frac{1+2\beta}{2-2\beta}(\frac{4\pi c^{2}}{r_{c}^{2}})^{\beta}H^{-2\beta-2}\right)= \frac{8\pi G}{3}\rho+\Lambda_2,
  \end{equation}
  where $\Lambda_2$ is the cosmological constant.
 And, a de-Sitter solution exists when $\beta<1$. The acceleration  heavily depends on the cutoff $r_{c}$ which is related to the extra dimension closely. Indeed, this is a kind of dark
  energy model which has been first proposed in Ref.\cite{turner}. This model could  be also  served
as an effective holographic dark energy model in Ref.\cite{gong09} as well. If gravity can be explained by holographic
principle and thermodynamics, then, the accelerating phase which is
induced by modified gravity has the origin of holographic principle
and thermodynamics.

\section{The $f(R)$ Correction}\label{sec6}
 As the simplest  case of high-order gravity,  the action of the so-called $f(R)$
gravity  is an arbitrary function of
curvature scalar $R$. When $f(R)=R$, the Einstein's general
relativity is recovered.
 Here, we try to discuss the $f(R)$ gravity from the view of  entropic force.  The area law of the entropy in
$f(R)$ gravity is changed to \cite{noe},
 \begin{equation}
 \label{S3}
S_3=f'(R)\frac{k_Bc^{3}}{4 G\hbar}A,
 \end{equation}
where $R$ is the curvature scalar, the prime means a derivative with
respect to $R$ and  the subscript ``$3$" notes   the $f(R)$
correction background. Correspondingly, based on the holographic principle, the number of bits $N_{3}$
is
 \begin{equation}
 \label{N3}
N_3=f'(R)\frac{Ac^{3}}{ G\hbar}.
 \end{equation}
 For  $f(R)$ gravity, however, recently
shown in Refs.\cite{fR,Eling:2006aw} that in order to derive the corresponding field equations, a treatment with
non-equilibrium thermodynamics of spacetime with extra entropy productions is needed in the unified first law.

\subsection{The Modified Newton's Force For The $f(R)$ Correction}
With this definition of the number of bits $N_3$, mainly combined with
the equipartition law of energy, one could easily
get that
 \begin{equation}
F_3=\frac{G}{f'(R)}\frac{M m}{ r^{2}}.
 \end{equation}
By defining $\tilde{f}(R)=f(R)-R$,  when $\tilde{f}'(R)\ll1$, the leading terms of the related  Taylor
expansion is obtained,
 \begin{equation}
F_3=\frac{GM m}{
r^{2}}\left(1+\tilde{f}'(R)+\tilde{f}^{'2}(R)\right).
 \end{equation}
If $f(R)=R$, where $\tilde{f}'(R)=0$,  Newton's gravity is  recovered. The constraints from Newton's law depend on the exact form of
$f(R)$ which require
$\tilde{f}'(R)\ll1$ in the small distances.

\subsection{The Friedmann Equation For The $f(R)$ Correction}
Using the corrected form  $N_3$ in $f(R)$ gravity, we get the change of the number of
bits,
 \begin{equation}
  dN_3=f'(R)\frac{ 8\pi c^{3}\tilde{r}_{A}}{ G \hbar}d
\tilde{r}_{A}+f''(R)\frac{4\pi c^{3} \tilde{r}_A^{2}}{
G\hbar}\frac{d R}{d\tilde{r}_{A}}d\tilde{r}_{A},
 \end{equation}
where  $R=-6(2H^{2}+\dot{H})$.
Based on the equipartition law of energy,  the change
of the total energy could be expressed as
 \begin{equation}
 \label{dE3}
dE_3=\frac{c^{4}f'(R)}{G}d\tilde{r}_A-c^{4}f''(R)\frac{H(24H\dot{
H}+6\ddot{H})}{2 G\dot{H}}d\tilde{r}_{A}.
 \end{equation}
 Meanwhile, if we treat the Misner-Sharp mass unchanged which is another definition of the energy,
 by using
Eqs. (\ref{dE2}) and (\ref{dE3}),
 the Raychaudhuri equation is obtained,
 \begin{equation}
 \label{fr1}
 \dot H\left(f'(R)-f''(R)\frac{H(24H\dot{ H}+6\ddot{H})}{2 \dot{H}}\right)=-4\pi G (\rho +p).
 \end{equation}
When $f(R)=R$, we obtain the standard Friedmann equation  by using
the energy conservation equation. Unfortunately, Eq.(\ref{fr1}) in not the correct Raychaudhuri
equation in $f(R)$ gravity. The reason may be traced back to that
the $f(R)$ gravity is a non-thermodynamical equilibrium process, so
the equipartition law of energy  may not be used in this way.   The description of the energy in the $f(R)$ correction may be not correct at all.

However, if we use the effective energy density and the pressure in
$f(R)$ gravity as below \cite{Bamba:2009id}:
 \begin{eqnarray}
 &&\tilde{\rho}= \frac{1}{8\pi G f'}\left(-\frac{(f-R f')}{2}-3Hf''\dot{R}\right)+\frac{\rho}{f'}, \\
 && \tilde{p}=\frac{1}{8\pi G f'}\left((f-R f')/2+f''\ddot{R}+f'''\dot{R}^{2}+6f''\dot{R}\right)+\frac{p}{f'}.
  \end{eqnarray}
 and keep the ``effective" entropy as
 $\tilde{S}=\frac{k_{B}c^{3}}{4G\hbar}A$, from the definitions (\ref{dE}) and (\ref{dE2}), one could  get
  \begin{equation}
\dot H f'(R)-\frac{H d f'(R)}{2dt}-\frac{d^{2}f'(R)}{d^{2}t}=-4\pi G
(\rho +p).
  \end{equation}
Combined with the energy density conserved equation of matter,   the
  Friedman equation corresponding to the action could also be obtained
  \begin{equation}
 H^{2}=\frac{8\pi
 G}{3f'(R)}\left[\rho+\frac{R f'(R)-f(R)}{2}-3H\dot{R}f''(R)\right].
  \end{equation}
 In a short summary, in $f(R)$ gravity, the entropic force scenario may not tell the exact thermodynamical process. A thermal equilibrium process exists or not in $f(R)$ gravity is still problem. Here, for the basis of the equipartition law of energy, indeed we treat this process as a thermal equilibrium state and we failed to get the ``correct'' Friedmann equation with the entropy (\ref{S3}). But if we choose the effective entropy as the Einstein gravity shown, the Friedmann equation corresponding to the action could be gotten.

\section{Conclusions}\label{sec7}
In this letter,  by considering three different corrections to the
horizon entropy, a trial research on the application of entropic
force  has been done in three different modified gravities.
The modifications of area law of entropy are the logarithmic
correction, the power-law correction and the $f(R)$ correction.
Following the new notion of gravity which are assumed as one kind of
entropic force, the modified law of Newton's gravity is obtained which
coincides with the existing observations
with suitable parameters.

To get the application of the entropic force in
cosmology, we try to get the Friedmann equations based on the notion
of entropic force. The energies defined by  the
equipartition law of energy and the Misner Sharp mass were used.
According to the form of entropy, we modified the
 form of the number of bits $N$, and the energy based on  the equipartition  law of energy is changed. By matching the energy given by
 the energy equipartition law and the Misner-Sharp mass,
  the modified Friedmann equations can be given out
  with the help of the energy density conserved equation of matter. Specifically, for the
logarithmic and power-law corrections to the entropy, the Friedmann
equation got an extra term (e.g. $H^\alpha$ term). The results suggest
that the accelerating phase in our universe has the possibility of being
 explained by entropic force. It may be an emergent phenomenon based on
 the holographic principle and thermodynamics. During the process of
 deriving the Friedmann equation, the partition law of energy is critical
  which should be used in a thermal equilibrium system. For the logarithmic
  and power-law corrections, we regarded that the energy equipartition law
  could be used. But, in the $f(R)$ correction,  by  using the equipartition
  law of energy, the choice of the entropy is still a problem.

However, there are still a lot of questions with the new
notion of gravity  of entropic force, our work
is just a pioneer which gives out an attempting study
to explain the Friedmann equation in the modified gravity.
Through the entropic force  is extended to the modified gravity
and the validity of the equipartition law of energy is checked
in our discussions, the limit of the equipartition law of energy
 is still worthy of further discussions.


\section*{Acknowledgements}
We thank Prof. Rong-gen Cai, Prof. Miao Li, Dr. Li-ming Cao, Prof.
Fu-wen Shu and Dr. Hong-bao Zhang for fruitful discussions, the
anonymous discussions on http://limiao.net/ as well. This work was supported by  China Postdoc Grant
No.20100470237£¬ the Ministry of Science and Technology of China
national basic science Program (973 Project) under grant Nos.
2007CB815401 and 2010CB833004, the National Natural Science
Foundation of China key project under grant Nos. 11005164, 10533010
and 10935013, and the Distinguished Young Scholar Grant 10825313,
and the Natural Science Foundation Project of CQ CSTC under grant
No. 2009BA4050 and 2010BB0408.

\providecommand{\newblock}{}

\end{document}